\documentstyle[prl,aps,psfig,multicol]{revtex}
\begin{document}
\draft
\title{Novel Phenomena in Dilute Electron Systems in Two Dimensions}
\author{M.~P.~Sarachik$^{(a)}$ and S.~V.~Kravchenko$^{(b)}$}
\address{$^{(a)}$Physics Department, City College of the City
University of New York, New York, New York 10031}
\address{$^{(b)}$Physics Department, Northeastern University,
Boston, Massachusetts 02115}
\maketitle
\makeatletter
\global\@specialpagefalse
\def\@oddhead{{\it \hfill To appear as a Perspective in the Proceedings of the National Academy of Sciences}}
\let\@evenhead\@oddhead
\makeatother
\begin{multicols}{2}
{\bf For the past two decades, all two-dimensional systems of electrons were
believed to be insulating in the limit of zero temperature.  We review recent experiments that provide evidence for an unexpected transition to a conducting phase at very low electron densities.  The nature of this phase is not understood, and is currently the focus of intense theoretical and experimental
attention.}\vspace{.3cm}

BACKGROUND

A two-dimensional (2D) system of electrons or ``holes'' (a hole, or missing
electron, behaves like a positively charged electron) is one in which
the positions of the electrons and their motion are restricted to a plane.
Physical realizations can be found in very thin films, sometimes at the
surface
of bulk materials, in ``quantum well'' systems such as GaAs/AlGaAs
that are specifically engineered for this purpose, and in the silicon
metal-oxide-semiconductor field-effect transistors described below.
Two-dimensional electron systems have been studied for nearly forty years
\cite{ando}, and
have yielded a number of important discoveries of physical
phenomena that directly reflect the quantum mechanical nature of our world.
These include the integer Quantum Hall Effect (QHE), which reflects the
quantization of electron states by a magnetic field, and the fractional
Quantum Hall Effect, which is a manifestation of the quantum mechanics of
many electrons acting together in a magnetic field to yield curious effects
like fractional (rather than whole) electron charges\cite{stormer}.

For nearly two decades it was believed that in the absence of an external
magnetic field ($H=0$) all two-dimensional systems of electrons are insulators
in the limit of zero temperature.  The true nature of the conduction was
expected to be revealed only at sufficiently low temperatures; in
materials such as highly conducting thin films, this was thought to require
unattainably low temperatures in the $\mu$Kelvin range.  Based on a scaling
theory for non-interacting electrons\cite{abrahams79}, these expectations
were further supported by theoretical work for
weakly interacting electrons \cite{altshuler80}.

Confirmation that two-dimensional systems of electrons are insulators in
zero field was provided by a beautiful series of experiments in thin metallic
films \cite{osheroff} and silicon metal-oxide-semiconductor field-effect
transistors\cite{bishop,pepper}, where the
conductivity was shown to display weak logarithmic corrections leading to
infinite resistivity in the limit of zero temperature.  It was therefore quite
surprising when recent experiments in silicon metal-oxide-semiconductor
field-effect transistors suggested that a transition from insulating to
conducting behavior occurs with increasing electron density at a very low
critical density, $n_c\sim10^{11}$~cm$^{-2}$\cite{kravchenko}.
These experiments were performed on unusually high quality samples,
allowing measurements at considerably lower electron densities than had
been possible in the past.  First viewed with considerable skepticism, the
finding was soon confirmed for silicon metal-oxide-semiconductor field-effect
transistors fabricated in
other laboratories\cite{popovic97}, and then for other
materials, including p-type SiGe structures \cite{coleridge97},
p-type GaAs/AlGaAs heterostructures \cite{hanein98,simmons98}, and
n-type AlAs \cite{papadakis98} and GaAs/AlGaAs heterostructures
\cite{hanein98a}.

It was soon realized that the low electron (and hole) densities at which
these observations were made correspond to a regime where the energy of the
repulsive Coulomb interactions between the electrons exceeds the Fermi energy
(roughly, their kinetic energy
of motion) by an order of magnitude or more.
For example, at an electron density $n_s=10^{11}$~cm$^{-2}$
in silicon metal-oxide-semiconductor field-effect transistors, the Coulomb
repulsion energy,
$U_c\sim e^2(\pi n_s)^{1/2}/\epsilon$, is about 10~meV while the Fermi
energy, $E_F=\pi n_s\hbar^2/2m^*$, is only 0.55~meV.  (Here $e$
is the electronic charge,
$\epsilon$ is the dielectric constant, and $m^*$ is the effective mass
of the
electron).  Rather than being a small perturbation, as has been generally
assumed in theoretical work done to date, interactions instead
provide the dominant energy in these very
dilute systems.\vspace{.3cm}

EXPERIMENTS

The inset to Fig.~1(a) shows a schematic diagram of the band structure of a
silicon metal-oxide-semiconductor field-effect transistor consisting of
a thin-film metallic gate deposited on an
oxide layer adjacent to lightly p-doped silicon, which serves as a source of
electrons.  A voltage applied between the gate and the oxide-silicon
interface causes the conduction and valence bands to bend, as shown in the
diagram, creating a potential minimum which traps electrons in a
two-dimensional layer perpendicular to the plane of the page.  The magnitude
of the applied voltage determines the degree of band-bending and thus the
depth of the potential well, allowing continuous control
of the number of electrons trapped in the two-dimensional system at the
interface.

For a very high-mobility (low disorder) silicon metal-oxide-semiconductor
field-effect transistor, the resistivity
is shown at several fixed temperatures as a
function of electron density in Fig.~1(a).  There is a well defined
crossing at a ``critical'' electron density, $n_c$, below which the
resistivity increases as the temperature is decreased, and above which the
reverse is true.  This can be seen more clearly in Fig.~1(b) where the
resistivity is plotted as a function of temperature for various fixed
electron densities.  A resistivity that increases with decreasing
temperature generally signals an approach to infinite resistance at $T=0$,
that is, to insulating behavior; a resistivity that decreases as the
temperature is lowered is characteristic of a metal if the resistivity
tends to a finite value, or a superconductor or perfect conductor if the
resistivity tends to zero.
The crossing point of Fig.~1(a) thus signals a transition from insulating
behavior below $n_s<n_c$ to conducting behavior at higher densities
($n_s>n_c$).  Similar behavior obtains in other materials at critical
densities determined by material parameters such as effective masses and
dielectric constants.  The value of the resistivity at the transition (the
``critical resistivity'' $\rho_c$) in all systems remains on the order of
$h/e^2$, the quantum unit of resistivity.

The electrons' spins play a crucial role in these low-density
materials, as demonstrated by their dramatic response to a magnetic field
applied parallel to the plane of the two-dimensional system.  We note that
an in-plane magnetic field couples only to the electron spins and does not
affect their orbital motion.  The parallel-field magnetoresistance is shown
for
a silicon metal-oxide-semiconductor field-effect transistor in Fig.~2 for
electron densities spanning the critical density $n_c$ at a temperature
of 0.3 K.  The
resistivity increases by more than an order of magnitude with increasing
field,
saturating to a new value in fields above $2$ or $3$ Tesla
\cite{simonian97a,pudalovH} above which the spins are presumably fully
aligned.  The total change in resistance is larger at lower temperatures and
for higher mobility samples, exceeding two or three order of magnitude in
some cases.  Although first thought to be associated only with the
suppression of the conducting phase, the fact that very similar
magnetoresistance is found for electron densities above and below the
zero-field critical density indicates that this is a more general feature of
dilute two-dimensional electron systems.\vspace{.3cm}

SOME OPEN QUESTIONS

Strongly interacting systems of electrons in two dimensions are currently the
focus
of intense interest, eliciting a spate of theoretical attempts to
account for the presence and nature of the unexpected conducting phase.
Most postulate esoteric new states of matter, such as a low-density conducting phase first considered by
Finkelshtein\cite{finkelshtein84}, a perfect metallic state
\cite{dobrosavljevic97}, non-Fermi liquid behavior \cite{chakravarty98},
and several types of superconductivity \cite{superconductivity}.

A number of relatively more mundane suggestions have been advanced that
attribute the unusual behavior seen in Figs.~1 and 2 to effects that are
essentially classical in nature.  These include a vapor/gas separation in the
electron system\cite{he}, temperature- and
field-dependent filling and emptying of charge traps unavoidably introduced
during device fabrication at the
oxide-silicon interface
\cite{maslov}, and temperature-dependent screening associated with
such charged traps \cite{dassarma}.  Although some may strongly
advocate
a particular view, all would agree that no consensus has been reached.

A great deal more experimental information will be required before the
behavior of these systems is understood.  Information will surely
be obtained in the near future from NMR, tunneling
studies, optical investigations, and other techniques.  One crucially
important question that needs to be resolved by experiment is the
ultimate fate of the resistivity in the conducting phase in the limit of
zero temperature.  Data in all 2D systems showing the unusual
metal-insulator transition indicate that, following the rapid (roughly
exponential) decrease with decreasing temperature shown in Fig.~1(b),
the resistivity levels off to a constant, or at most weakly
temperature-dependent, value.  The temperature at which this leveling
off occurs decreases, however, as the transition is
approached\cite{kapitulnik}.
The question is whether the
resistivity of dilute two-dimensional systems tends to a finite value or
zero in the zero-temperature limit as the transition is approached.  If the
resistivity remains finite, this would rule out
superconductivity\cite{superconductivity}
or perfect conductivity\cite{dobrosavljevic97}.
The question may then revert to whether localization of the electrons
reasserts itself at very low temperatures, yielding an
insulator as originally expected.  There are
well-known experimental difficulties associated with cooling the
electron system to the same temperature as the lattice and bath (that is,
the temperature measured by the thermometer), and these experiments
will require great skill, care and patience.

An equally important issue is the magnetic response of the electron system.
Superconductors expel magnetic flux and are strongly diamagnetic, while
Finkelshtein's low-density phase would give a strongly paramagnetic signal.
There are very few electrons in a low-density, millimeter-sized,
$100 \AA$-thick layer, and measurements of the magnetization will be
exceedingly difficult.

In closing, we address a crucial question regarding the nature of the
apparent, unexpected zero-field metal-insulator transition: do these
experiments signal the presence of unanticipated phases and new
phenomena in strongly interacting two-dimensional electron systems, or can
the observations be explained by invoking classical effects such as recharging
of traps in the oxide or temperature-dependent screening?  Some recent
experiments suggest the former.  A Princeton-Weizmann
collaboration\cite{hanein99} has demonstrated that the magnetic field-induced
phase transition between integer Quantum Hall Liquid and insulator
(the QHE-I transition)
evolves
smoothly and continuously to the metal-insulator transition in zero magnetic
field discussed in this paper, raising the possibility that the two
transitions
are closely
related.  This conjecture is supported by the strong similarity between the
temperature dependence of the resistivity in zero magnetic field and in
the Quantum Hall Liquid phase\cite{krav99}.  Additional insight can be
obtained from a comparison of the ``critical'' resistivity, $\rho_c$, at the
zero-field metal-insulator transition and the critical resistivity,
$\rho_{QHE-I}$, at the QHE-I transition measured for the same sample
\cite{hanein99,krav99,coleridge99}.  Fig.~3(a) shows values of
the zero-field critical resistivity, $\rho_c$, for a number of samples of
different 2D electron and hole systems: $\rho_c$ varies by an order of magnitude,
between approximately $10^4$ and $10^5$~Ohm, and exhibits no apparent
systematic behavior.  In contrast, Fig.~3(b) shows that the ratio
$\rho_c/\rho_{QHE-I}$ is close to unity when measured on the same sample for
three different materials.  Since the QHE-I transition is clearly a
quantum phase transition, this suggests that the zero-field metal-insulator transition is a
quantum phase transition as well.  The intriguing relationship between
critical
resistivities for these two transitions shown for only a very few samples in
Fig.~3(b) clearly needs further confirmation.  Future work will surely resolve
whether, and what, exciting and unanticipated physics is required to account
for the puzzling and fascinating recent observations in two
dimensions.\vspace{.3cm}

AKNOWLEDGMENTS

We are grateful to P.~T.~Coleridge for sharing his data \cite{coleridge99}
with us prior to publication.  M.~P.\ Sarachik thanks the US Department of
Energy for support under grant No.\ DE-FG02-84-ER45153.  M.~P.~S.\ and
S.~V.~Kravchenko acknowledge support by NSF grant DMR~98-03440.

\newpage
\end{multicols}
FIGURE CAPTIONS\vspace{.5cm}

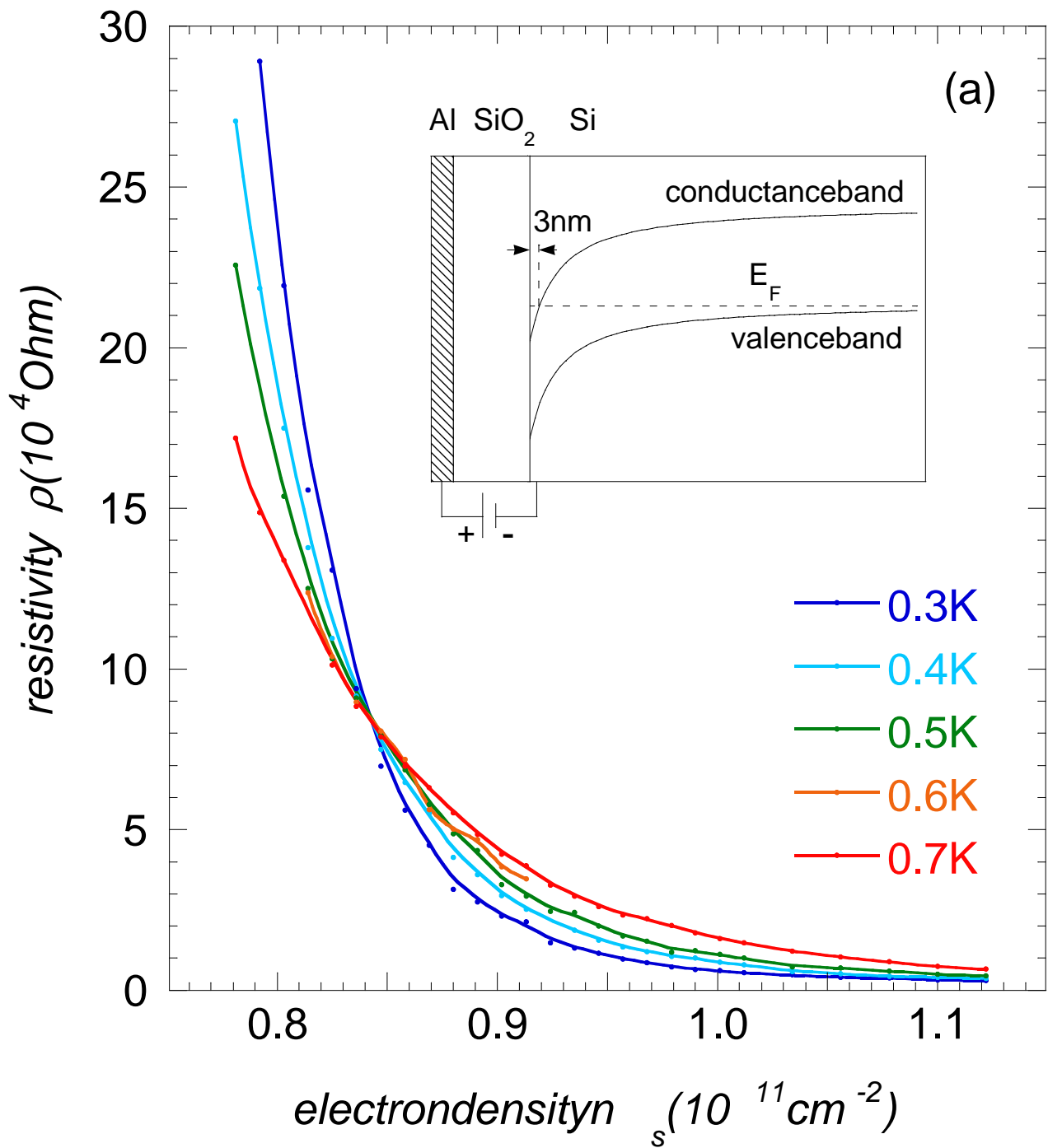
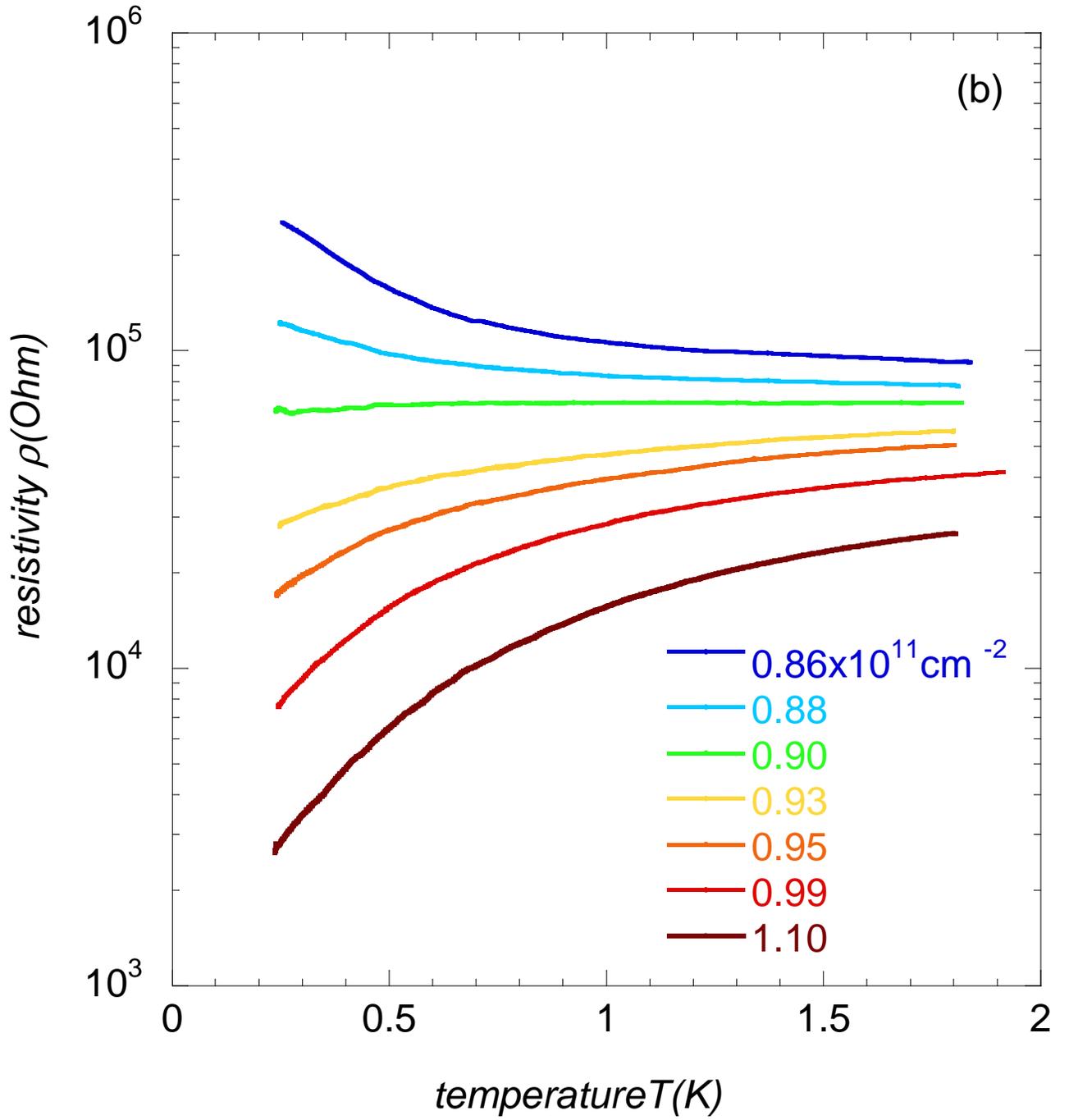
Figure 1:

(a): Resistivity as a function of electron density for the
two-dimensional system of electrons in a high-mobility silicon
metal-oxide-semiconductor field-effect transistor.  The different curves
correspond to different temperatures.  Note that at low densities the
resistivity increases with decreasing temperature (insulating behavior),
while the reverse is true for higher densities (conducting behavior).
The inset shows a schematic diagram of the electron bands to illustrate how
a two-dimensional layer is obtained (see text).

(b):  Resistivity as a function of temperature for the two-dimensional
system of electrons in a silicon MOSFET.  Different curves are for different electron densities.\vspace{.5cm}

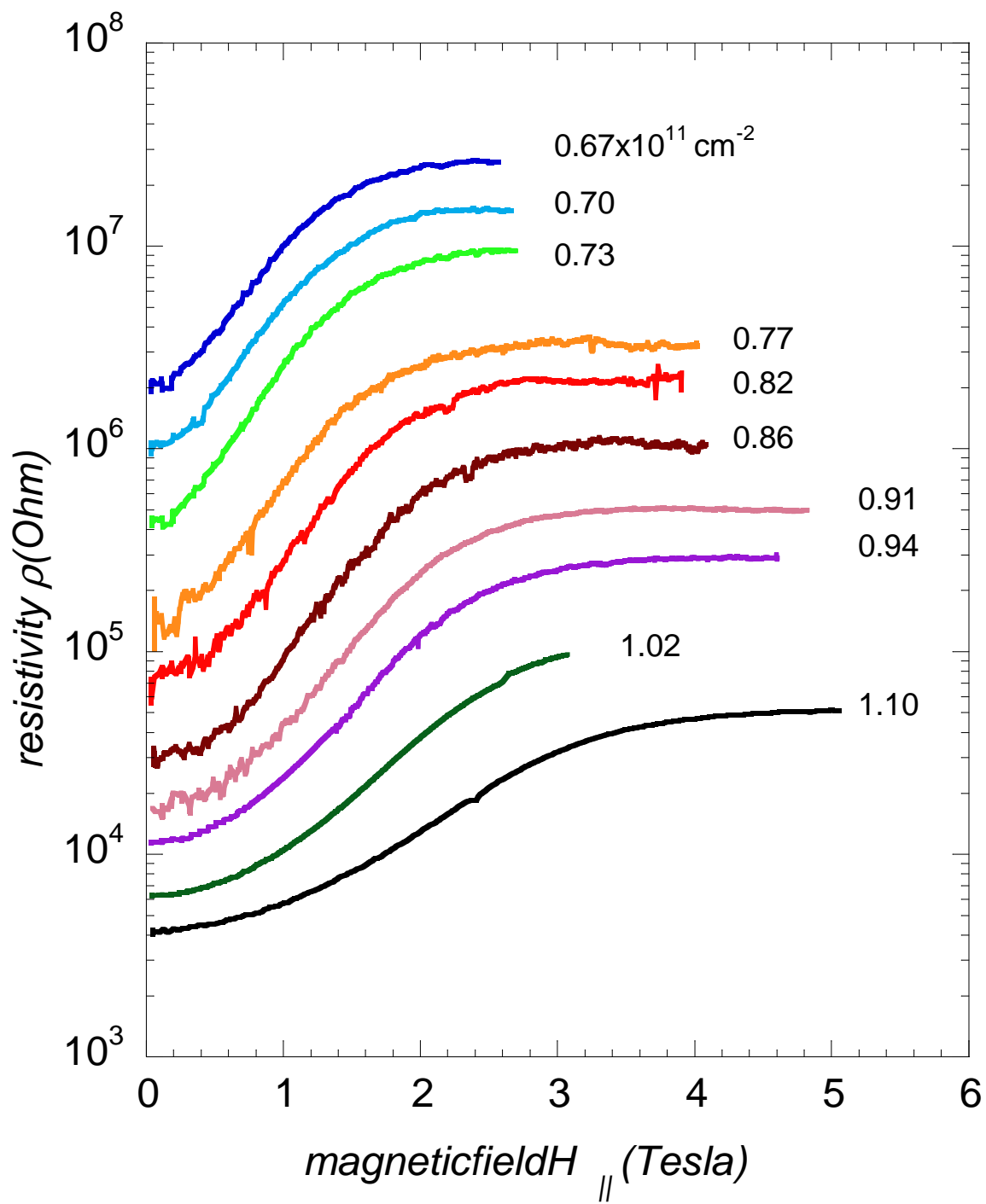
Figure 2:

For different electron densities, the resistivity at $0.3$~Kelvin is plotted
as a function of magnetic field applied parallel to the plane of the
two-dimensional system of electrons in a silicon MOSFET.  The top 
three curves are
insulating while the lower curves are conducting in the absence of a magnetic
field.  The response to parallel field is qualitatively the same in the two
phases, varying continuously across the transition.\vspace{.5cm}

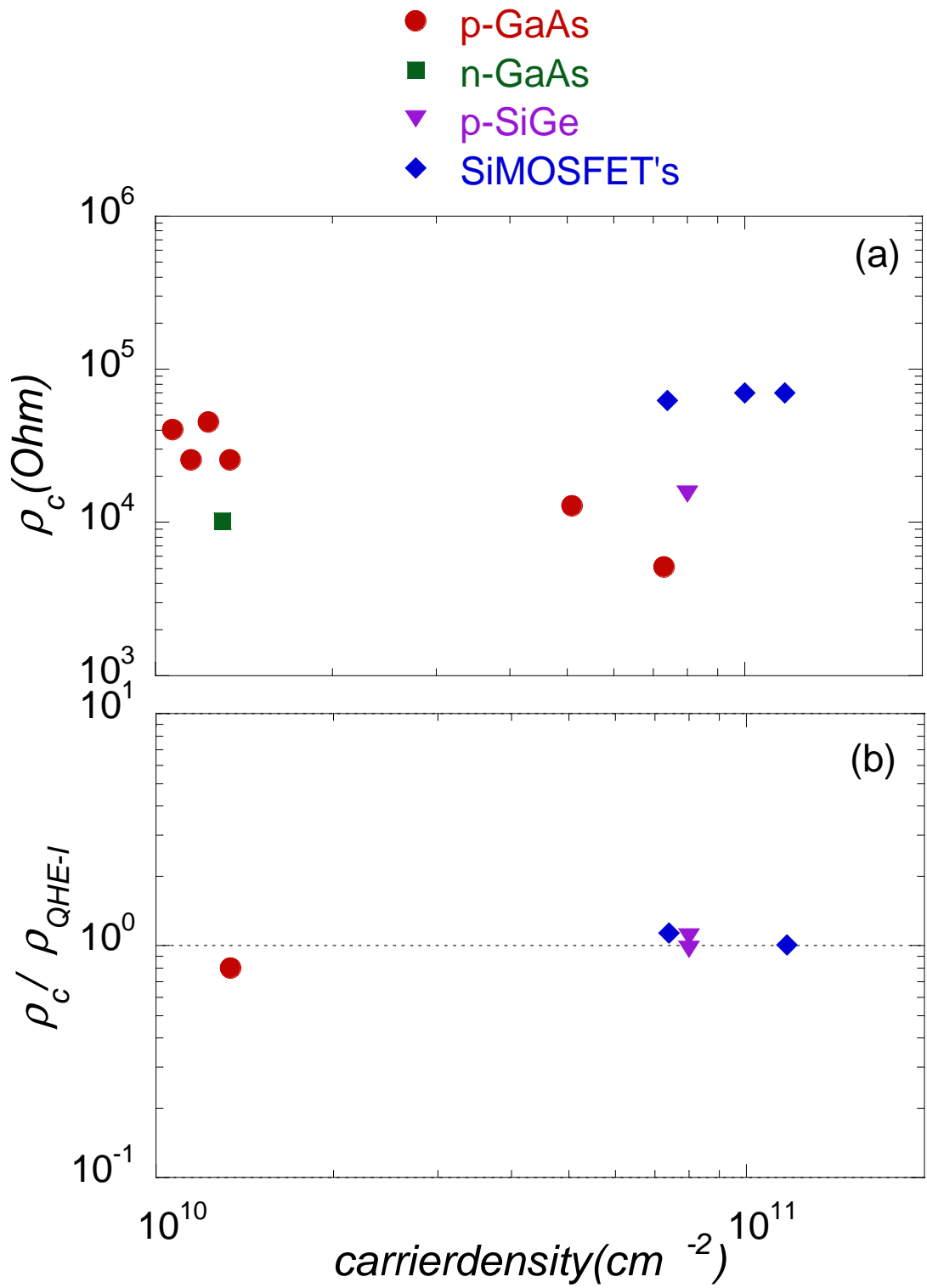
Figure 3:

(a):  The critical resistivity, $\rho_c$, which separates the conducting and
insulating phases in zero magnetic field is shown for several 2D
systems for which the transition occurs at different electron (or hole)
densities, shown along the $x$-axis.  Although the critical resistivity is of
the order of the quantum unit of resistivity, $h/e^2\approx26$~kOhm, it varies
by about a factor of 10.\\

(b):  For several materials, measurements of $\rho_c$ and
$\rho_{QHE-I}$ on the {\it same} sample yield ratios $\rho_c/\rho_{QHE-I}$
that are near unity.  Here $\rho_c$
is the critical resistivity separating the conducting and insulating phases in
the absence of magnetic field and $\rho_{QHE-I}$ is the critical resistivity
at the transition from the Quantum Hall Liquid to the insulator in finite
magnetic field.  Data were obtained for p-GaAs/AlGaAs heterostructures from
Refs.~[11,12,25], for n-GaAs/AlGaAs from Ref.~[14], and for p-SiGe from  Refs.~[10,27].
\end{document}